\pdfoutput=1
\documentclass[aip,apl,twocolumn,reprint,preprintnumbers,amsmath,amssymb]{revtex4-1}
\usepackage{graphicx}
\usepackage{amsmath}
\usepackage{mathcomp}
\usepackage{dcolumn}
\usepackage{bm}
\begin{document}

\title{Cooling electrons from 1 K to 400 mK with V-based  nanorefrigerators}

\author{O. Quaranta}
\email{orlando.quaranta@sns.it}
\author{P. Spathis}
\author{F. Beltram}
\author{F. Giazotto}
\email{giazotto@sns.it}
\affiliation{NEST, Istituto Nanoscienze-CNR and Scuola Normale Superiore, Piazza S. Silvestro 12, I-56127 Pisa, Italy}

\begin{abstract}
The fabrication and operation of V-based superconducting nanorefrigerators is reported. Specifically, electrons in an Al island are cooled thanks to hot-quasiparticle extraction provided by tunnel-coupled V electrodes.
Electronic temperature reduction down to 400 mK starting from 1 K is demonstrated with a cooling power $\sim 20$ pW at $1$ K for a junction area of $0.3\,\mu$m$^2$. 
The present architecture extends to higher temperatures refrigeration based on tunneling between superconductors
and paves the way to the implementation of a multi-stage on-chip cooling scheme operating from  above $1$ K down to the mK regime.

\end{abstract}

\maketitle 

The investigation of heat-transport at the nanoscale is currently the focus of an intense research effort \cite{giazottoRMP}.
In this context, solid-state refrigeration with an emphasis on superconducting tunnel structures is under the spotlight \cite{giazottoRMP,nahum1994,leivo1996,pekola2004,tirelli2008,kafanov2009,manninen1999,clark2004}. 
These superconducting refrigerators yielded
the demonstration of sizable electron-temperature reductions \cite{leivo1996,pekola2004,clark2004} 
allowing to reach few tens of mK in optimized structures. 
These structures make a widespread use of Al as superconductor since this material makes it possible to fabricate high-quality tunnel junctions. On the other hand, however, Al limits the operation temperature of the coolers 
to a few hundreds of mK owing to its reduced critical temperature.  
Indeed the exploitation of a superconducting gap larger than that of Al is technologically challenging but would give the opportunity
to extend this cooling method up to temperatures around or above 1K.

Here we report the fabrication and operation of V/AlO$_x$/Al all-superconducting refrigerators which demonstrate up to $60\%$ reduction of electron temperature when operated at 1K, and provide an estimated cooling-power density of $\sim 65$ pW/$\mu$m$^2$. 
The ease of fabrication and the performance demonstrated here indicate the V-Al material combination as an ideal candidate for the implementation of on-chip cascaded refrigerators operating from above $\sim1$ K down to the mK regime. 

\begin{figure}[t]
\includegraphics[width=7.7cm]{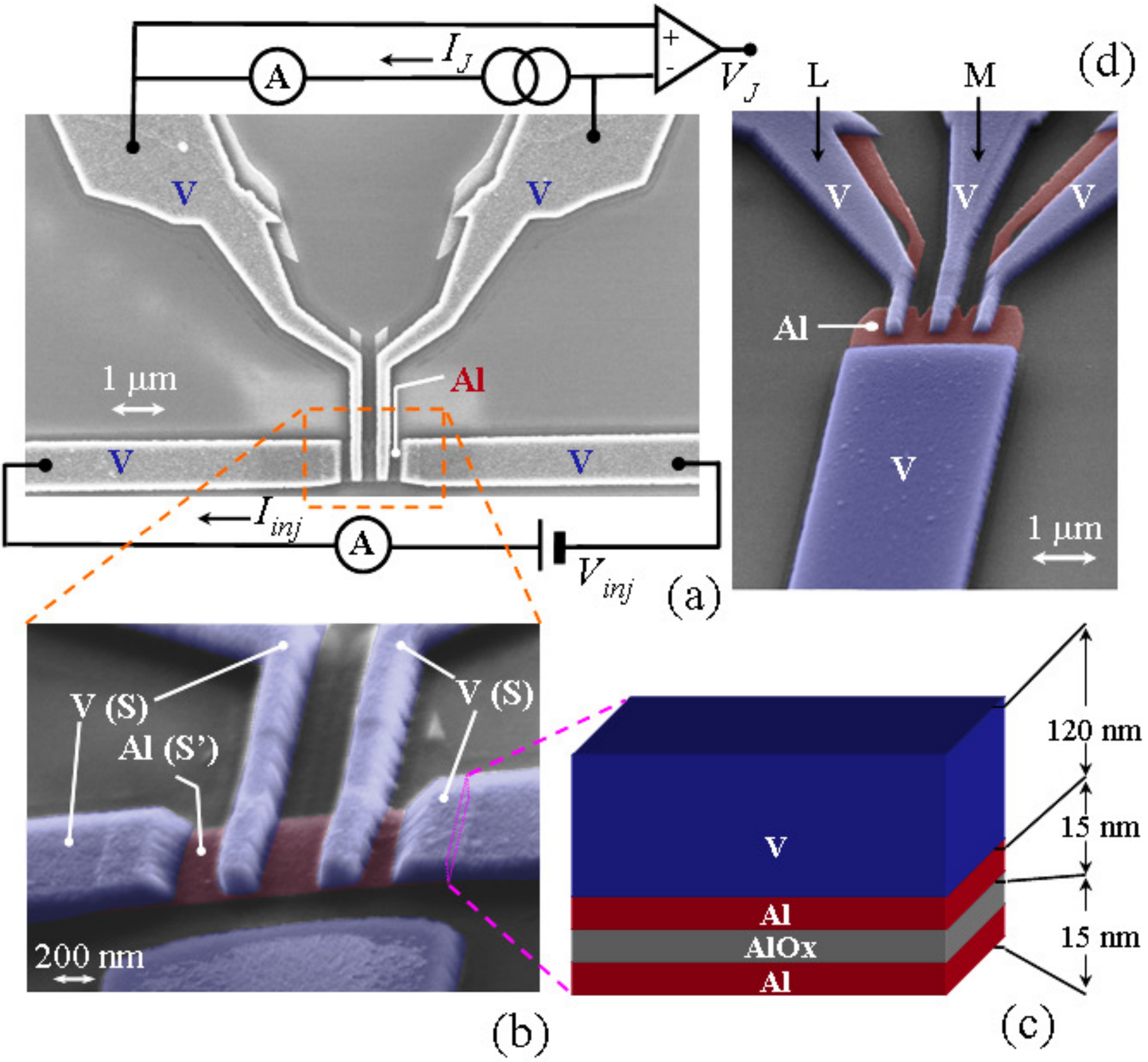}
\caption{\label{fig1}(a) Scanning electron micrograph of a typical V-based nanorefrigerator along with a scheme of the measurement setup.
$I_{inj}$ is the current driven through the cooling junctions while $I_J$ is that flowing through the Josephson thermometers.  
(b) A pseudo-color blow-up of the device core. V is the larger gap superconductor (S) acting as electron refrigerator.
(c) Cross-section of the junctions (not to scale). (d) Pseudo-color  micrograph of a reference structure (see text).}
\end{figure}

Figure~\ref{fig1}(a) shows a scanning electron micrograph of a V-based double refrigerator along with a scheme of the measurement setup. 
A blow-up of the device which consists of a 15-nm-thick 600$\times$1200 nm$^2$ Al island (S') tunnel-coupled to four V electrodes (S) is displayed in Fig.~\ref{fig1}(b).
Samples were fabricated by e-beam lithography and two-angle shadow-mask metal evaporation.
Figure~\ref{fig1}(c) shows a cross-section of the junctions. 
The barriers were obtained by oxidation of the Al layer in 100 mTorr of O$_2$ for 20' leading to tunnel junctions with $\sim 300\,\Omega \mu\text{m}^2$ specific resistance.
Direct evaporation of V onto AlO$_x$ degrades junction quality, therefore a 15-nm-thick Al interlayer was first deposited to prevent direct contact between V and the oxide layer.
The resulting junctions showed excellent stability against thermal cycling and reduced aging. 
The two lateral junctions (with normal-state resistance $R_T\simeq 2$ k$\Omega$ each) are operated as electron coolers through current injection $I_{inj}$,\cite{giazottoRMP,tirelli2008} whereas the two inner (with $R_T\simeq 2-3$ k$\Omega$ each) are used as Josephson thermometers to probe electron temperature in the island. 
The critical temperature of the Al island was $T_c^{S'}\simeq 1.55$ K whereas that of the 120/15-nm-thick V/Al bilayer was $T_c^{S}\simeq 4$~K.
Figure~\ref{fig1}(d) shows a reference structure that will be discussed later in the text. It consists of a few V tunnel junctions connected to a large-volume Al electrode. 

Samples were characterized down to 280 mK in a filtered $^3$He cryostat as follows.
We first calibrated the Josephson thermometers \cite{giazottoRMP,tirelli2008}, i.e., the series connection of the two central SIS' junctions (I denotes an insulator). 
To this end the dependence of the equilibrium Josephson critical current ($I_c$) on bath temperature ($T_{bath}$) was determined. These results were then used to convert the supercurrent amplitude under injection $V_{inj}$ into an electronic temperature value in the Al island. 
In the following we shall present results obtained in two representative devices referred to as A and B. 
\begin{figure}[t!]
\includegraphics[width=7.7cm]{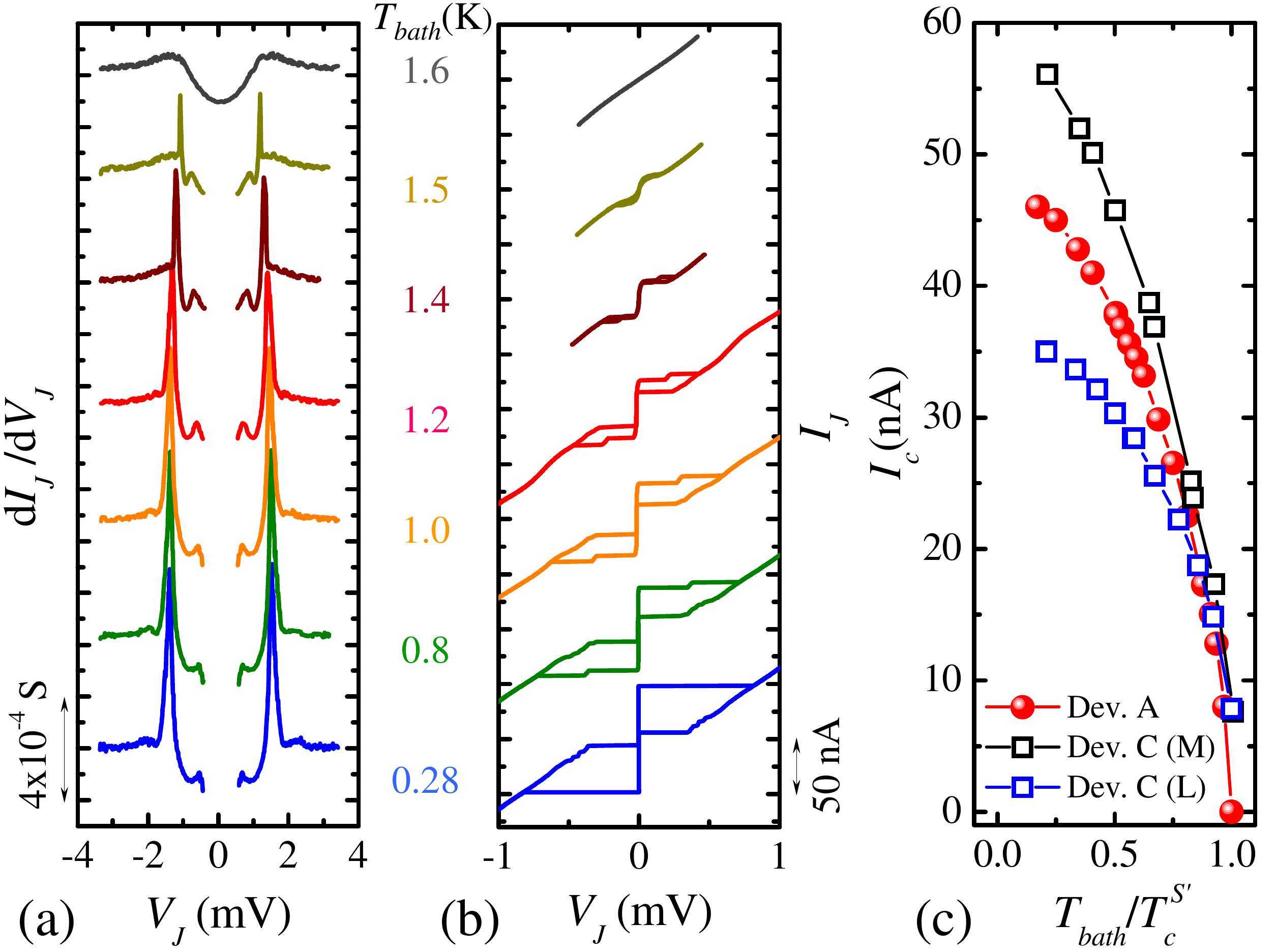}
\caption{\label{fig2}(a) $dI_J/dV_J$ vs $V_J$ and (b) $I_J$ vs $V_J$ characteristics of thermometer junctions in device A at different  $T_{bath}$. In (a) and (b) curves are vertically offset for clarity. 
(c) Equilibrium critical current $I_c$ vs reduced temperature for three different junctions. Solid dots refer to device A, whereas open squares refer to the middle (M) and left (L) junctions in device C [see Fig. ~\ref{fig1}(d)].
}
\end{figure}

Figure~\ref{fig2}(a) shows the differential conductance $dI_J/dV_J$ vs. voltage $V_J$ for thermometer junctions in device A at different $T_{bath}$ values. 
Curves are vertically offset for clarity.
The two peaks appearing at higher voltage correspond to $V_{J}=\pm 2(\Delta_S+\Delta_{S'})/e$, while those appearing at lower energy to $V_{J}=\pm 2(\Delta_S-\Delta_{S'})/e$. This is the typical behavior of the tunneling characteristic between two different superconductors at finite temperature \cite{Tinkham}. Here $\Delta_S$ ($\Delta_{S'}$) is the energy gap of V/Al (Al), and $e$ is the electron charge.
As the temperature is raised the two peaks merge, coinciding at temperatures larger than $T_c^{S'}$.
The temperature evolution of the peaks closely follows the Bardeen-Cooper-Schrieffer (BCS) behavior \cite{Tinkham}. 
From the low-temperature voltage position of the peaks we deduced 
 $\Delta_S\simeq 580~\mu$eV and $\Delta_{S'}\simeq 200~\mu$eV.

The corresponding $I_J$ vs $V_J$ characteristics are displayed in Fig. \ref{fig2}(b).
Junctions exhibit a hysteretic dc Josephson effect, as expected for underdamped tunnel weak links \cite{Tinkham}. 
At intermediate temperatures a finite-bias step is observed which stems from a slight asymmetry between the junctions which
leads to a non simultaneous switching to the resistive state.  
The Josephson coupling persists up to $\simeq 1.55$ K. Above this temperature the island is driven into the normal state.

\begin{figure}[t!] 
\includegraphics[width=7.7cm,keepaspectratio]{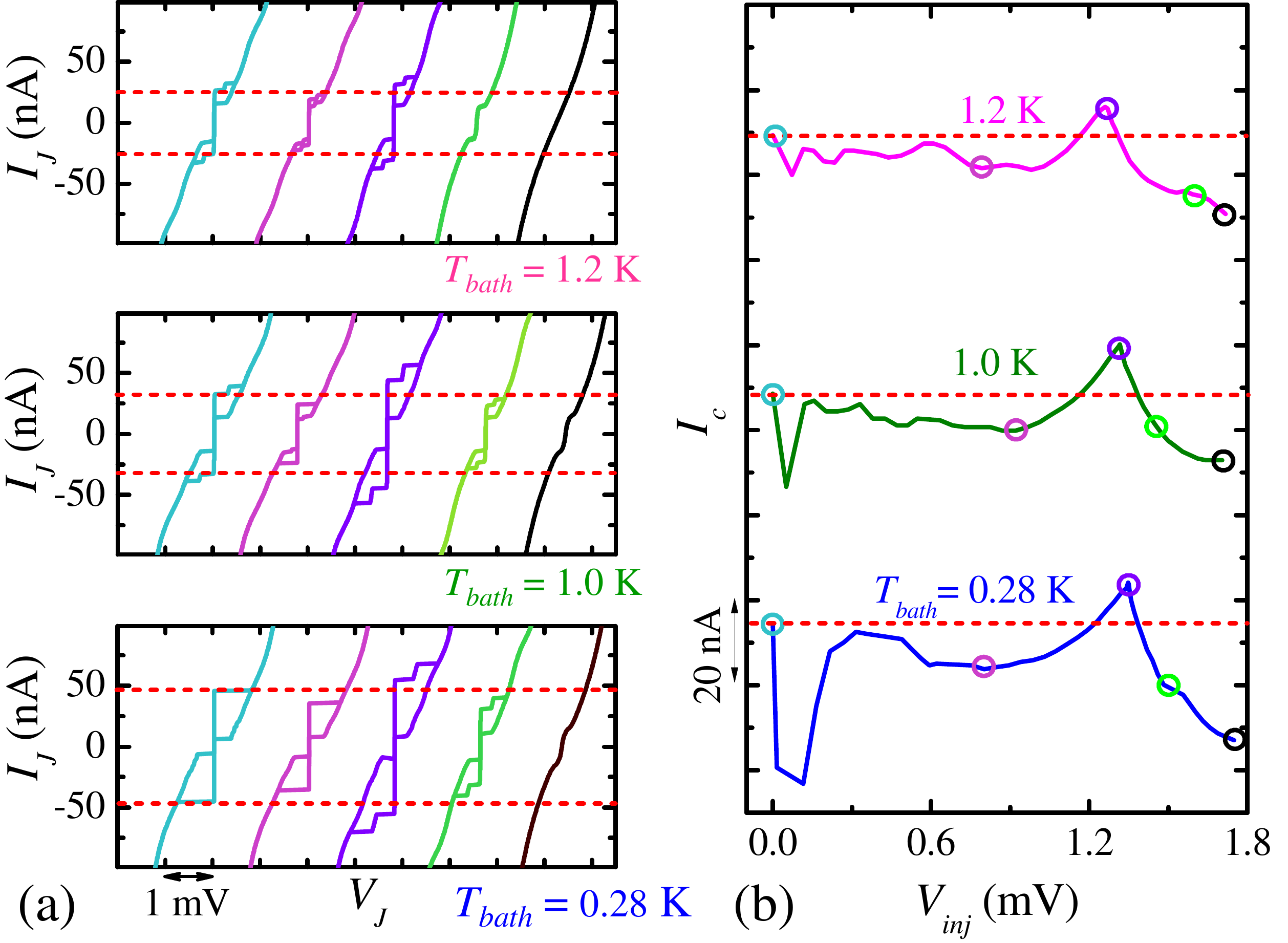}
\caption{\label{fig3}(a) $I_J$ vs $V_J$ characteristics of sample A thermometer junctions under several increasing $V_{inj}$ at three different $T_{bath}$. The curves are horizontally offset for clarity. 
(b) Maximum critical current $I_c$ vs $V_{inj}$ at the same $T_{bath}$. Colored circles indicate $V_{inj}$ values under which the characteristics shown in panel (a) were measured. The curves are vertically offset for clarity.
Dashed lines in (a) and (b) indicate the equilibrium critical current at each corresponding temperature.}
\end{figure}

The full temperature evolution of the equilibrium supercurrent at $V_J=0$ for the same junctions is shown in Fig. \ref{fig2}(c) (solid dots). In particular, $I_c$ monotonically increases by reducing $T_{bath}$ without showing saturation even at the lowest temperatures.
This behavior markedly differs from the Ambegaokar-Baratoff prediction \cite{AB} and is expected for tunnel junctions embedding an interlayer like the present \cite{golubov1995,brammertz2001}. 
In order to validate our temperature calibration and rule out spurious overheating in the small island we also measured the $I_c(T_{bath})$ dependence on similar junctions fabricated on a large-volume Al electrode, i.e., device C shown in Fig. \ref{fig1}(d) (open squares). 
Besides showing different $I_c$, the junctions exhibit the same temperature dependence when normalized at each corresponding maximum supercurrent
thus confirming the role of the Al interlayer.

Figure~\ref{fig3}(a) shows the $I_{J}$ vs $V_{J}$ characteristics for thermometers in sample A with increasing values of $V_{inj}$. Curves are horizontally offset for clarity; each panel refers to a different $T_{bath}$. 
The evolution of the Josephson supercurrent upon increasing  $V_{inj}$ is non-monotonic: at all measured temperatures
$I_c$ is initially suppressed, then it increases and even exceeds the equilibrium value. This latter behavior occurs thanks to hot-quasiparticle extraction provided by V reservoirs which \emph{cools} the electron population in the island \cite{giazottoRMP,tirelli2008}.
Further increase of $V_{inj}$ leads to a decay of $I_c$ then followed by suppression of the supercurrent for larger injection voltage.

The full evolution of $I_c$ under $V_{inj}$ is shown in Fig. \ref{fig3}(b) for the same $T_{bath}$ values.
Colored circles indicate the $V_{inj}$ values at which the characteristics shown in Fig.~\ref{fig3}(a) were measured. 
Supercurrent enhancement is clearly observable even at $1.2$ K. 
Such a nonmonotonic behavior of the Josephson current closely resembles that reported and extensively discussed in Ref. [4], 
and originates from the presence of a floating superconductor coupled to biased superconducting electrodes. 

\begin{figure}[t!]
\includegraphics[width=7.7cm,keepaspectratio]{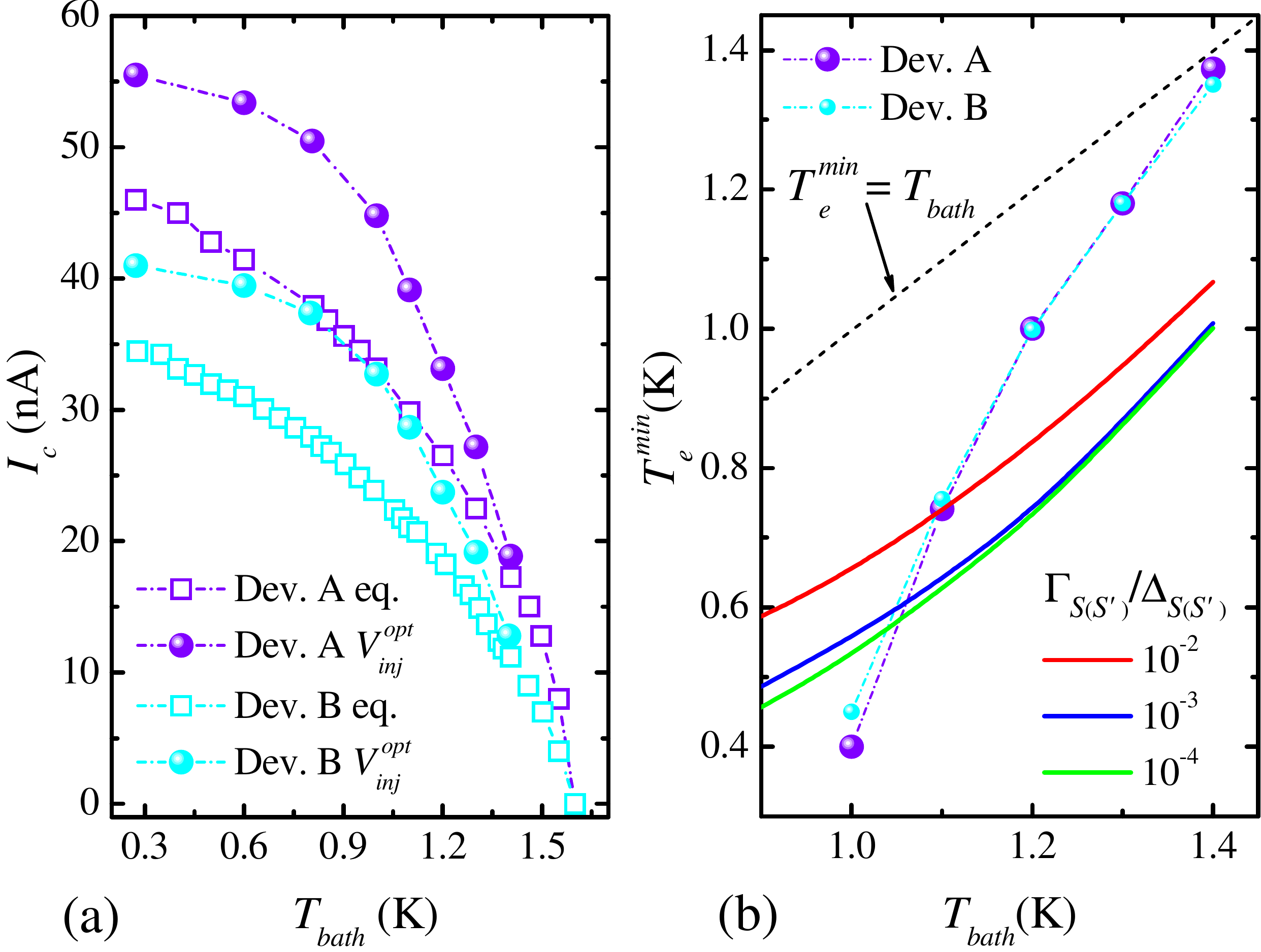}
\caption{\label{fig4}
(a) $I_c$ vs $T_{bath}$ for device A and B at the equilibrium, i.e., at $V_{inj}=0$ (open squares), and at $V_{inj}^{opt}$ (solid dots). (b) Minimum electron temperature $T_e^{min}$ at $V_{inj}^{opt}$ vs $T_{bath}$ for both samples  (solid dots).
Dash-dotted lines are guides to the eye. 
Full lines are the prediction for an ideal SIS'IS refrigerator with the same parameters as in our samples calculated for different $\Gamma_{S(S')}$ values. Dashed line represents $T_e^{min}=T_{bath}$.   }
\end{figure}

Figure \ref{fig4}(a) shows $I_c$ vs $T_{bath}$ (open squares) for both samples. 
As anticipated [see also Fig. 2(c)], $I_c$ increases monotonically upon lowering $T_{bath}$, and at $280$ mK exhibits values as large as $\sim 35$ nA and $\sim 46$ nA for sample A and B, respectively. 
For comparison, the maximum $I_c$ at the optimized injection voltage $V_{inj}^{opt}$, i.e., that which maximizes the supercurrent at each $T_{bath}$, is also displayed (solid dots).
Below $\simeq 1.3$ K the out-of-equilibrium supercurrent exceeds that at equilibrium, attaining $\sim 42$ nA and $\sim 56$ nA  in sample A and B, respectively, at 280 mK. 

The minimum electron temperature $T_e^{min}$ achieved under injection in the island at different $T_{bath}$ values is shown in Fig. \ref{fig4}(b) (solid dots). $T_e^{min}$ was determined by comparing the maximum $I_c$ at $V_{inj}^{opt}$ with the equilibrium values [panel (a)].  A dramatic cooling is achieved at 1 K where $T_e^{min}$ values as low as $\sim 400$ mK and $\sim 450$ mK in sample A and B, respectively, were obtained which correspond to up to $60\%$ temperature reduction.
Moreover, substantial cooling is obtained even at $1.2$ K where $T_e^{min}\sim 1$ K. At lower $T_{bath}$ we cannot determine $T_e^{min}$ owing to the lack of a temperature calibration below 280 mK.

Insight into the operation of the refrigerator can be gained based on 
a simple model which includes energy relaxation in the island and the heat flux originating from injection. 
Although rather idealized, this analysis allows us to grasp the essential features of our system.
Under bias voltage $V_{inj}$  the heat current flowing from S' to the S electrodes is given by \cite{giazottoRMP,tirelli2008} $\dot{Q}=2(e^2R_T)^{-1}\int{d\varepsilon}\varepsilon\mathcal{N}_S(\tilde{\varepsilon})\mathcal{N}_{S'}(\varepsilon)[f_0(\varepsilon,T_e^{S'})-f_0(\tilde{\varepsilon},T_e^{S})],$
where $\tilde{\varepsilon}=\varepsilon-eV_{inj}/2$, $f_0(\varepsilon,T)$ is the Fermi-Dirac function at temperature $T$,  
and $\mathcal{N}_{S,S'}(\varepsilon)=|\text{Re}[(\varepsilon+i\Gamma_{S,S'})/\sqrt{(\varepsilon+i\Gamma_{S,S'})^2-\Delta^2_{S,S'}(T_e^{S,S'})}]|$ is the smeared BCS density of states where $\Gamma_{S,S'}$ accounts for nonidealities in S(S'). \cite{giazottoRMP,pekola2004}
We consider a transport regime where strong quasiparticle interaction in S' forces the electron system to retain a thermal \emph{quasiequilibrium} at temperature $T_e^{S'}$ differing in general from $T_{bath}$.
The actual $T_e^{S'}$ upon current injection is the result of the competition between $\dot{Q}$ and other relaxation mechanisms transferring energy into S'. 
At low $T_{bath}$ the predominant contribution comes from electron-phonon interaction $\dot{Q}_{e-ph}^{S'}$ [see Eq. (2) of Ref. [13]] which allows energy exchange between electrons and lattice phonons \cite{giazottoRMP,timofeev2009}. 
The steady-state $T_e^{S'}$ is obtained by solving the energy-balance equation $\dot{Q}(V_{inj},T_{bath},T_e^{S'})+\dot{Q}_{e-ph}^{S'}(T_{bath},T_e^{S'})=0$. 

Solid lines in Fig. \ref{fig4}(b) represent $T_e^{min}$ as obtained from the solution of the balance equation for different values of $\Gamma_{S(S')}$.
In particular an increase of $\Gamma_{S(S')}$ leads for each $T_{bath}$ to a reduced cooling which stems from enhanced heating originating from subgap current in tunnel junctions \cite{giazottoRMP,pekola2004,rajauria2008}. 
At higher $T_{bath}$ values the experimental data deviate from the calculations which predict a lower $T_e^{min}$,  whereas the agreement is improved for $T_{bath}\leq 1.1$ K. 
The nature of our junctions which include an (Al) interlayer  might explain the deviation from this rather simplified model that actually holds for ideal SIS' junctions.

We should finally comment on the available cooling power. Based on the expression for $\dot{Q}$ we estimate that our structures provide a maximum $\dot{Q}$ of $\sim 20$ pW at $1$ K corresponding to $\sim 65$ pW/$\mu$m$^2$ areal cooling-power density. 
This value can be increased by either lowering the junctions specific resistance or by optimizing the thickness of the Al interlayer.

In conclusion, we have reported the fabrication and operation of V-based nanorefrigerators. These coolers show a significant temperature reduction, i.e.,  up to $60\%$ when operating at 1 K, and provide an areal cooling-power density of $\sim 65$ pW/$\mu$m$^2$. 
Furthermore, the fabrication protocol yields junctions with excellent stability and reduced aging.
Our results pave the way to the implementation of a solid-state multi-stage refrigeration platform which would enable on-chip cooling from the base temperature of a $^4$He cryostat down to the mK regime.   

We acknowledge  the NanoSciERA project ``NanoFridge'' for partial financial support.

\end{document}